# The importance of transparency and reproducibility in artificial intelligence research

Benjamin Haibe-Kains[1,2,3,4,5,$], George Alexandru Adam[3,5], Ahmed Hosny[6,7], Farnoosh Khodakarami[1,2], MAQC Society Board[8,*], Levi Waldron[9], Bo Wang[2,5,10], Chris McIntosh[2,5,10], Anshul Kundaje[11], Casey S. Greene[12,13], Michael M. Hoffman[1,2,3] Jeffrey T. Leek[14], Wolfgang Huber[15], Alvis Brazma[16], Joelle Pineau[17,18], Robert Tibshirani[19,20], Trevor Hastie[19,20], John P.A. Ioannidis[19,20,21,22], John Quackenbush[24,25,26], Hugo J.W.L. Aerts[6,7,27,20]

**Affiliations**
[1] Princess Margaret Cancer Centre, University Health Network, Toronto, Ontario, Canada
[2] Department of Medical Biophysics, University of Toronto, Toronto, Ontario, Canada
[3] Department of Computer Science, University of Toronto, Toronto, Ontario, Canada
[4] Ontario Institute for Cancer Research, Toronto, Ontario, Canada
[5] Vector Institute for Artificial Intelligence, Toronto, Ontario, Canada
[6] Artificial Intelligence in Medicine (AIM) Program, Brigham and Women's Hospital, Harvard Medical School, Boston, MA, USA
[7] Radiation Oncology and Radiology, Dana-Farber Cancer Institute, Brigham and Women's
[8] Massive Analysis Quality Control (MAQC) Society Board of Directors
[9] Department of Epidemiology and Biostatistics and Institute for Implementation Science in Population Health, CUNY Graduate School of Public Health and Health Policy, New York, NY, USA
[10] Peter Munk Cardiac Centre, University Health Network, Toronto, Ontario, Canada
Hospital, Harvard Medical School, Boston, MA, USA
[11] Department of Genetics, Stanford University School of Medicine, Stanford, CA, USA
[12] Dept. of Systems Pharmacology and Translational Therapeutics, Perelman School of Medicine, University of Pennsylvania, Philadelphia, PA, USA
[13] Childhood Cancer Data Lab, Alex's Lemonade Stand Foundation, Philadelphia, PA, USA
[14] Department of Biostatistics, Johns Hopkins Bloomberg School of Public Health, Baltimore MD, USA
[15] European Molecular Biology Laboratory, Genome Biology Unit, Heidelberg, Germany
[16] European Molecular Biology Laboratory, European Bioinformatics Institute, EMBL-EBI, Hinxton, UK
[17] McGill University, Montreal, QC, Canada
[18] Montreal Institute for Learning Algorithms, QC, Canada
[19] Meta-Research Innovation Center at Stanford (METRICS), Stanford, CA, USA
[20] Department of Biomedical Data Science, Stanford University School of Medicine, Stanford, CA, USA
[21] Departments of Medicine, of Health Research and Policy, and of Biomedical Data Science, Stanford University School of Medicine, Stanford, CA, USA

[22] Department of Statistics, Stanford University School of Humanities and Sciences, Stanford, CA, USA
[23] Department of Epidemiology and Population Health, Stanford University School of Medicine, Stanford, CA, USA
[24] Department of Biostatistics, Harvard T.H Chan School of Public Health, Boston, MA, USA
[25] Channing Division of Network Medicine, Brigham and Women's Hospital, Boston, MA, USA
[26] Department of Data Science, Dana-Farber Cancer Institute, Boston, MA, USA
[27] Radiology and Nuclear Medicine, Maastricht University, Maastricht, Netherlands
[28] Cardiovascular Imaging Research Center, Massachusetts General Hospital, Harvard Medical School, Boston, MA, USA

**[$] Corresponding Author**
Benjamin Haibe-Kains: bhaibeka@uhnresearch.ca

**[*] Massive Analysis Quality Control (MAQC) Society Board of Directors**
Thakkar Shraddha, Rebecca Kusko, Susanna-Assunta Sansone, Weida Tong, Russ D. Wolfinger, Christopher Mason, Wendell Jones, Joaquin Dopazo, Cesare Furlanello

**Abstract:**
In their study, McKinney et al. showed the high potential of artificial intelligence for breast cancer screening. However, the lack of detailed methods and computer code undermines its scientific value. We identify obstacles hindering transparent and reproducible AI research as faced by McKinney et al and provide solutions with implications for the broader field.

**Main Text:**
The evaluation of deep learning for the detection of breast cancer from mammograms by McKinney and colleagues[1] showed promising improvements in screening performance, while highlighting challenges around the reproducibility and transparency of artificial intelligence (AI) research. They assert that their system improves the speed and robustness of breast cancer screening, generalizes to populations beyond those used for training, and outperforms radiologists in specific settings. Upon successful prospective validation, this new system holds great potential for streamlining clinical workflows, reducing false positives, and improving patient outcomes. However, the absence of sufficiently documented methods and computer code underlying the study effectively undermines its scientific value. This shortcoming limits the evidence required for others to prospectively validate and clinically implement such technologies. Here, we identify obstacles hindering transparent and reproducible AI research as faced by McKinney et al. and provide potential solutions with implications for the broader field.

Scientific progress depends upon the ability of independent researchers to (1) scrutinize the results of a research study, (2) reproduce the study's main results using its materials, and (3) build upon them in future studies[2]. Publication of insufficiently documented research violates the core principles underlying scientific discovery[3,4]. The authors state *"The code used for training the models has a large number of dependencies on internal tooling, infrastructure and hardware, and its release is therefore not feasible"*. Computational reproducibility is indispensable for robust AI applications[5,6] more complex methods demand greater transparency[7]. In the absence of code, reproducibility falls back on replicating methods from textual description. Although, the authors claim that *"all experiments and implementation details are described in sufficient detail in the Supplementary Methods section to support replication with non-proprietary libraries"*, key details about their analysis are lacking. Even with sufficient description, reproducing complex computational pipelines based purely on text is a subjective and challenging task[8,9].

More specifically, the authors' description of the model development as well as data processing and training pipelines lacks critical details. The definition of multiple hyperparameters for the model's architecture (composed of three networks referred to as the Breast, Lesion, and Case models) is missing (Table 1). The authors did not disclose the parameters used for data augmentation; the transformations used are stochastic and can significantly affect model performance[10]. Details of the training pipeline were also missing. For instance, they state that the mini-batches were sampled to contain an equal proportion of negative and positive examples, potentially leading to multiple instances of the same patients in a given epoch. Deep learning optimization algorithms such as stochastic gradient descent typically operate under the assumption that a given sample is provided to the model exactly once per epoch. The lack of detail regarding the per-batch balancing of classes prevents replicating the training pipeline.

There exist numerous frameworks and platforms to make artificial intelligence research more transparent and reproducible (Table 2). For the sharing of code, these include Bitbucket, GitHub, and GitLab among others. The multiple software dependencies of large-scale machine learning applications require appropriate control of software environment, which can be achieved through package managers including Conda, as well as container and virtualization systems, including Code Ocean, Gigantum and Colaboratory. If virtualization of the Mckinney et al. internal tooling proved to be difficult, they could have released the computer code and documentation. The authors could also have created toy examples to show how new data must be processed to generate predictions. As for the trained model, many platforms allow sharing of deep learning models, including TensorFlow Hub, ModelHub.ai, ModelDepot, and Model Zoo with support for multiple frameworks such as PyTorch, Caffe, and TensorFlow. Since the authors created their model with the publicly available TensorFlow library, sharing of the model should be trivial. In addition to improving accessibility and transparency, such tools can significantly accelerate model development, validation, and transition into production and clinical implementation.

Another crucial aspect of ensuring reproducibility lies in access to the data the models were derived from. In their study, McKinney et al. used two large datasets under license, properly disclosing this limitation in their publication. Sharing of patient health information is highly regulated due to privacy concerns. Despite these challenges, sharing of raw data has become more common in biomedical literature, increasing from under 1% in the early 2000s to 20% today[11]. However, if the data cannot be shared, the model predictions and data labels themselves should be released, allowing further statistical analyses.

Although sharing of code and data is widely seen as a crucial part of scientific research, the adoption varies across fields. In fields such as genomics, complex computational pipelines and sensitive datasets have been shared for decades[12]. Guidelines related to genomic data are clear, detailed, and most importantly, enforced. It is generally accepted that all code and data are released alongside a publication. In other fields of medicine and science as a whole, this is much less common, and data and code are rarely made available. For scientific efforts where a clinical application is envisioned and human lives would be at stake, we argue that the bar of transparency should be set even higher. If data cannot be shared with the entire scientific community, because of licensing or other insurmountable issues, at a minimum a mechanism should be set so that some highly-trained, independent investigators can access the data and verify the analyses. This would allow a truly adequate peer-review of the study and its evidence before moving into clinical implementation.

Many egregious failures of science were due to lack of public access to code and data used in the discovery process[13,14]. These unfortunate lessons should not be lost on either journal editors or its readers. Journals have an obligation to hold authors to the standards of reproducibility that benefit not only other researchers, but also the creators of a given method. Making one's methods reproducible may surface biases or shortcomings to authors before publication[15]. Preventing external validation of a model will likely reduce its impact and could lead to unintended consequences[15]. The failure of McKinney et al. to share key materials and information transforms their work from a scientific publication open to verification into a promotion of a closed technology.

We have high hopes for the utility of AI methods in medicine. Ensuring that these methods meet their potential, however, requires that these studies be reproducible. Unfortunately, the biomedical literature is littered with studies that have failed the test of reproducibility, and many of these can be tied to methodologies and experimental practices that could not be investigated due to failure to fully disclose software and data. This is even more important for applications intended for use in the diagnosis or treatment of human disease.

**Competing Interests**
AH is a shareholder of and receives consulting fees from Altis Labs. MMH received a GPU Grant from Nvidia. HJWLA is a shareholder of and receives consulting fees from Onc.AI. BHK is a scientific advisor for Altis Labs. GAA, FK, LW, BW, CM, AK, CSG, JTL, WH, AB, JP, RT, TH, JPAI and JQ declare no other competing interests related to the manuscript.

**Author Contributions**
BHK and GAA wrote the first draft of the manuscript. BHK and HJWLA designed and supervised the study. AH, FK, LW, BW, CM, AK, CSG, MMH, JTL, WH, AB, JP, RT, TH, JPAI and JQ contributed to the writing of the manuscript.

**Table 1: Essential hyperparameters for reproducing the study for each of the three models (Lesion, Breast, and Case), including those missing from the description in Mckinney et al.**

|  | Lesion | Breast | Case |
|---|---|---|---|
| Learning rate | Missing | 0.0001 | Missing |
| Learning rate schedule | Missing | Stated | Missing |
| Optimizer | Stochastic gradient descent with momentum | Adam | Missing |
| Momentum | Missing | Not applicable | Not applicable |
| Batch size | 4 | Unclear | 2 |
| Epochs | Missing | 120,000 | Missing |

**Table 2: Frameworks and platforms to share code, software dependencies and deep learning models to make artificial intelligence research more transparent and reproducible.**

| Resource | | URL |
|---|---|---|
| *Code* | | |
| BitBucket | | https://bitbucket.org |
| GitHub | | https://github.com |
| GitLab | | https://about.gitlab.com |
| *Software dependencies* | | |
| Conda | | https://conda.io |
| Code Ocean | | https://codeocean.com |
| Gigantum | | https://gigantum.com |
| Colaboratory | | https://colab.research.google.com |
| *Deep learning models* | | |
| TensorFlow Hub | | https://www.tensorflow.org/hub |
| ModelHub | | http://modelhub.ai |
| ModelDepot | | https://modeldepot.io |
| Model Zoo | | https://modelzoo.co |
| *Deep learning frameworks* | | |
| TensorFlow | | https://www.tensorflow.org/ |
| Caffe | | https://caffe.berkeleyvision.org/ |
| PyTorch | | https://pytorch.org/ |

| Author Name | ORCID |
|---|---|
| Benjamin Haibe-Kains | 0000-0002-7684-0079 |
| George-Alexandru Adam | 0000-0001-9084-3703 |
| Ahmed Hosny | 0000-0002-1844-481X |
| Levi Waldron | 0000-0003-2725-0694 |
| Bo Wang | /0000-0002-9620-3413 |
| Chris McIntosh | 0000-0003-1371-1250 |
| Anshul Kundaje | 0000-0003-3084-2287 |
| Casey S. Greene | 0000-0001-8713-9213 |
| Michael M. Hoffman | 0000-0002-4517-1562 |
| Jeffrey T. Leek | 0000-0002-2873-2671 |
| Wolfgang Huber | 0000-0002-0474-2218 |
| Alvis Brazma | 0000-0001-5988-7409 |
| Joelle Pineau | 0000-0003-0747-7250 |
| Robert Tibshirani | 0000-0003-0553-5090 |
| Trevor Hastie | 0000-0002-0164-3142 |
| John P. A. Ioannidis | 0000-0003-3118-6859 |
| John Quackenbush | 0000-0002-2702-5879 |
| Hugo JWL Aerts | 0000-0002-2122-2003 |

| MAQC Society Board Members | ORCID |
|---|---|
| Thakkar Shraddha | 0000-0002-2920-7713 |
| Rebecca Kusko | 0000-0001-5331-5119 |
| Susanna-Assunta Sansone | 0000-0001-5306-5690 |
| Weida Tong | 0000-0003-3488-6148 |
| Russ D. Wolfinger | 0000-0001-8575-0537 |
| Christopher Mason | 0000-0002-1850-1642 |
| Wendell Jones | 0000-0002-9676-5387 |
| Joaquin Dopazo | 0000-0003-3318-120X |
| Cesare Furlanello | 0000-0002-5384-3605 |